\documentclass[12pt]{article}
\usepackage{amsmath}
\usepackage{amsfonts}

\begin{document}

\newcommand{\bi}[1]{ \boldsymbol{#1} }
\newcommand{\rmi}[0]{\mathrm{i}}
\newcommand{\rme}[0]{\mathrm{e}}
\newcommand{\rmd}[0]{\mathrm{d}}
\newcommand{\varkappa}[0]{\kappa}
\newcommand{\Or}[0]{\mathrm{O}}
\newcommand{\JPA}[0]{\textit{J. Phys. A: Math. Gen.}}
\newcommand{\JPAT}[0]{\textit{J. Phys. A: Math. Theor.}}
\newcommand{\JMP}[0]{\textit{J. Math. Phys.}}
\newcommand{\PR}[0]{\textit{Phys. Rev.}}

\title{Wavelet-based integral representation for solutions of the wave equation}

\author{Maria V Perel and Mikhail S Sidorenko}

\maketitle \noindent
{\footnotesize\center  Department of Mathematical Physics, Physics Faculty, \\
St.Petersburg University, \\
Ulyanovskaya 1-1, Petrodvorets, St.Petersburg, 198904, Russia \\
\texttt{{mailto: perel@mph.phys.spbu.ru, M-Sidorenko@yandex.ru} }
\\}

\indent

\begin{abstract}
An integral representation of solutions of the wave equation as a superposition of other solutions of this equation is built. The solutions from a wide class can be used as building blocks for the representation. Considerations are based on mathematical techniques of continuous wavelet analysis. The formulas obtained are justified from the point of view of distribution theory. A comparison of the results with those by G. Kaiser is carried out. Methods of obtaining physical wavelets are discussed.
\end{abstract}
\vspace{20pt}
\newpage

\section{Introduction}
\indent

The aim of the paper is to find a new exact integral representation
of solutions of the wave equation. We consider a homogeneous
equation with constant coefficients in a three-dimensional space. We
represent a solution as a superposition of some other elementary
solutions of the wave equation, which can be taken from a wide
class. Although we use the term 'elementary', the solutions may be
rather complicated. We call them '\textit{elementary}', because we
represent other solutions as their superposition.

A well-known exact integral representation for the wave equation of
such a kind is the Fourier integral, where solutions are decomposed
into the superposition of plane waves. However it is sometimes
convenient to have elementary solutions well-localized in space.
This is useful for studying local properties of solutions and, among
other things, for studying the propagation of singularities. An
approximate integral representation of solutions as a superposition
of localized Gaussian beams is developed in
\cite{Popov}-\cite{Heyman-Felsen}. However, it is heuristic and
inexact. Other mathematical methods for developing an exact
representation of this kind are necessary.

In our paper, we use mathematical techniques of wavelet analysis. It
has been developed in the 80s of the 20th century. The main ideas of
wavelet analysis take their origin in group representation theory
and in the theory of coherent states (see \cite{Klauder, Perelomov}
and references therein). The first papers on continuous wavelet
analysis theory were motivated by applications to seismic wave
propagation \cite{Grossmann1, Grossmann}. These papers stimulated
interest in wavelet analysis. Nowadays a great many of books and
articles on wavelet analysis are available (see
\cite{Daubechies}-\cite{Kaiser-Book} for instance).

Continuous wavelet analysis provides  a reconstruction formula,
which allows us to represent functions as a superposition of a
family of "elementary" functions obtained from one function in a
special way. The techniques of continuous wavelet analysis imply the
following steps. First, we need to fix the Hilbert space of
functions under consideration. Second, we choose a certain function
called a 'mother wavelet' in this Hilbert space. The function must
satisfy a special admissibility condition. Next we specify a group
of transformations, which will be applied to the mother wavelet in
order to construct the family of wavelets. In this way, an
overcomplete set of functions is obtained. The group must have some
special properties (see \cite{Grossmann, Antoine-Book}). Next a
wavelet transform of an arbitrary function from the Hilbert space
under consideration is obtained as the scalar product of the
function and each of the wavelets from the family constructed. The
wavelet transform depends on group parameters and contains
information about local properties of the function (see
\cite{Antoine-Book}).

The wavelet transforms of all functions from the considered Hilbert
space themselves form a subspace of the Hilbert space of functions
of group parameters. The map from the space of functions to the
space of wavelet transforms is an isometry. This allows us to obtain
a reconstruction formula. We follow this scheme in our argument
below.

A special case of continuous wavelet analysis based on the analytic
signal transform in a three-dimensional space was first applied to
the wave equation by G. Kaiser in \cite{Kaiser-Book} and developed
in \cite{Kaiser2}-\cite{Kaiser4}. He obtained an integral
representation formula for solutions of the homogeneous wave
equation as a superposition of elementary solutions derived from one
fixed mother wavelet only, named by him the '\textit{physical
wavelet}'.

In outline, the content of the paper is as follows. First we present
in Section \ref{S2} a brief discussion of the idea of our method and
illustrate it with a few simple facts. A more detailed development
of the method is a subject of other sections. The aim of Section
\ref{S3} is to construct a wavelet-based integral representation for
solutions of the wave equation. We split the whole space of
solutions into a direct sum of two subspaces containing solutions
with positive and negative frequencies. In each of them we obtain a
family of elementary solutions by applying transformations to a
certain solution at a fixed moment of time. This solution must
satisfy an admissibility condition. The transformations are spatial
translations, dilations, and rotations, which are common for
standard wavelet analysis. We also apply a dilation of time. We
suggest to use the wavelet transform of the solution as coefficients
for its integral representation. We also show that the coefficients
can be expressed in terms of the solution itself and its
time-derivative at a fixed moment of time, and this does not require
the decomposition of the solution into positive- and
negative-frequency parts. A justification of the results from the
point of view of the theory of distributions is given. A detailed
comparison with the results of Kaiser has been carried out in
\cite{Perel-Sidorenko-2}, and a brief review of it is given here.

In Section \ref{S4}, we discuss the possibility of obtaining new
physical wavelets by means of some known methods of constructing
explicit exact solutions of the wave equation. We also consider
exponentially localized physical wavelets found and generalized in
\cite{Kiselev-Perel-OptSp, Kiselev-Perel-JMP, Perel-Fialkovsky},
wavelet properties of which has been studied in
\cite{Perel-Sidorenko-JPA}.

\section{Preliminary discussion \label{S2}}

Fourier analysis allows one to construct a solution $u(\bi{r}, t)$
of the wave equation
\begin{equation}
u_{tt}(\bi{r}, t) - c^2[u_{xx}(\bi{r}, t) + u_{yy}(\bi{r}, t)]=0,
\qquad c = \mathrm{const},
\end{equation}
as the superposition of plane waves. For example, a solution with
positive frequencies has the decomposition
\begin{eqnarray}\label{c-fourier}
u(\bi{r}, t) = \frac{1}{(2\pi)^2}\int\limits_{\mathbb{R}^2}
\rmd^2\bi{k} \, \widehat{u}(\bi{k}) \, \exp(\rmi \bi{k} \cdot \bi{r}
- \rmi |\bi{k}| c  t).
\end{eqnarray}
Here $\widehat{u}(\bi{k})$ is the Fourier transform of the function
$u(\bi{r}, 0)$. For the sake of simplicity, here we consider the
two-dimensional case. The aim of our work is to construct another
integral representation of a solution of the wave equation as the
superposition of its localized solutions from a wide class. Instead
of using Fourier analysis, we invoke mathematical techniques of
continuous wavelet analysis. First we consider a special case of
wavelet analysis to compare it with the Fourier transform. It allows
us to represent an arbitrary functions in terms different from
harmonic exponents. In the case of the Special choice of the mother wavelet, the representation of the
function $u(\bi{r})$ can be written in the form
\begin{eqnarray}\label{w01}
 u(\bi{r})=\frac{1}{C_{\varphi}} \int\limits_{\mathbb{R}^2}
\rmd^2 \bi{k} \int\limits_{\mathbb{R}^2} \rmd^2 \bi{b} \,\,
U(\bi{k}, \bi{b}) \, \exp\left[ - \frac{|\bi{k}|^2}{2\sigma^2}
|\bi{r} - \bi{b}|^2 \right] \left( \rme^{\rmi \bi{k} \cdot (\bi{r} -
\bi{b})} - \rme^{-\sigma^2/2} \right),
\end{eqnarray}
where $U(\bi{k}, \bi{b})$ is defined by the formula
\begin{equation}\label{w02}
U(\bi{k}, \bi{b}) \equiv |\bi{k}|^2 \int\limits_{\mathbb{R}^2}
\rmd^2 \bi{r} \, u(\bi{r}) \, \exp\left[
-\frac{|\bi{k}|^2}{2\sigma^2} \, |\bi{r} - \bi{b}|^2 \right] \left(
\rme^{-\rmi \bi{k} \cdot (\bi{r} - \bi{b})} - \rme^{-\sigma^2/2}
\right),
\end{equation}
and $C_{\varphi}$ is a constant that plays the role of $(2\pi)^2$ in
the Fourier formula (\ref{c-fourier}). If $\sigma \gg 1$, the second
term can be neglected, and this formula can be interpreted as the
Fourier transform of the function $u(\bi{r})$ multiplied by a
cutting exponent in the vicinity of the point $\bi{b}$. The larger $|\bi{k}|$ the smaller
the width of the cutting function.

Now we introduce a solution $\varphi^{\bi{k},\bi{b}}(\bi{r}, t)$ of
the wave equation, which is satisfies the following initial conditions
\begin{equation}\label{init-prel}
\left. \varphi^{\bi{k},\bi{b}}(\bi{r}, t) \right|_{t=0} = \exp\left[
- \frac{|\bi{k}|^2}{2\sigma^2} |\bi{r} - \bi{b}|^2 \right] \left(
\rme^{\rmi \bi{k} \cdot (\bi{r} - \bi{b})} - \rme^{-\sigma^2/2}
\right),
\end{equation}
\begin{equation}
\left. \frac{\partial}{\partial t} \varphi^{\bi{k},\bi{b}}(\bi{r},
t)\right|_{t = 0} = 0.\nonumber
\end{equation}
Then taking into account formula (\ref{w01}), we can obtain a
solution $\widetilde{u}(\bi{r}, t)$ of the wave equation as follows:
\begin{equation}
\widetilde{u}(\bi{r}, t)=\frac{1}{C_{\varphi}}
\int\limits_{\mathbb{R}^2} \rmd^2 \bi{k} \int\limits_{\mathbb{R}^2}
\rmd^2 \bi{b} \,\, U(\bi{k}, \bi{b})
\,\varphi^{\bi{k},\bi{b}}(\bi{r}, t),
\end{equation}
where $U(\bi{k}, \bi{b})$ is defined by the formula (\ref{w02}) and
\begin{equation}
\widetilde{u}(\bi{r}, t) |_{t=0} = u(\bi{r}).
\end{equation}
However we cannot find an explicit solution
$\varphi^{\bi{k},\bi{b}}(\bi{r}, t)$ satisfying (\ref{init-prel}).
There exists an exact highly localized solution of the wave equation
named the 'Gaussian wave packet', which was found in
\cite{Kiselev-Perel-JMP} and studied in \cite{Perel-Sidorenko-JPA}
and which has the following explicit form:
\begin{equation}\label{gwp1}
\varphi(\bi{r}, t) = \frac{1}{\sqrt{x + ct - \rmi\varepsilon}}
\exp\left(-p\sqrt{1 - \frac{\rmi\theta}{\varepsilon}}\right),
\,\,\,\, \theta = x - ct + \frac{y^2}{x + ct - \rmi\varepsilon}.
\end{equation}
It was shown that if $p \gg 1$, this solution is actually an
exponentially localized wave packet moving along the $OX$ axis with
speed $c$. Our aim is to construct an integral representation that
expresses solutions in terms of the Gaussian wave packet
(\ref{gwp1}) and other localized solutions of the wave equation.

The formulas (\ref{w01}) - (\ref{w02}) are formulas of continuous
wavelet analysis but written in the nonstandard notation. The
notation common for wavelet analysis is as follows. The spatial
frequency parameter $\bi{k}$ is viewed as $\bi{k} = (\cos\beta,
\sin\beta )/a$ and $a, \beta$ are taken as parameters. The formula
for a wavelet family reads
\begin{equation}
\varphi^{a, \beta, \bi{b}}(\bi{r}) = \frac{1}{a}
\varphi\left(\mathrm{M}^{-1}_{\beta}\, \frac{\bi{r} - \bi{b}}{a}
\right), \qquad \mathrm{M}_{\beta} = \left(\begin{array}{cc}
\cos\beta & -\sin\beta  \\
\sin\beta & \cos\beta
\end{array} \right),
\end{equation}
where $a$ defines a dilation, $\bi{b}$ defines a translation, and
$\beta$ defines a rotation of the argument of the function
$\varphi$. Continuous wavelet analysis allows one to represent an
arbitrary square-integrable function $u$ as the superposition of
wavelets $\varphi$:
\begin{equation}
u(\bi{r}) = \frac{1}{C_{\varphi}} \int\limits_{0}^{2\pi} \rmd \beta
\int\limits_{0}^{+\infty} \frac{\rmd a}{a^3} \,
\int\limits_{\mathbb{R}^2} \rmd^2\bi{b} \, U(a, \beta, \bi{b})
\varphi^{a, \beta, \bi{b}}(\bi{r}),
\end{equation}
\begin{equation}
\qquad C_{\varphi} \equiv \int\limits_{\mathbb{R}^2} \rmd^2
\bi{k}\frac{|\widehat{\varphi}(\bi{k})|^2}{|\bi{k}|^2},\nonumber
\end{equation}
where the coefficients of the decomposition are defined by the
formula
\begin{equation}
U(a, \beta, \bi{b}) \equiv \int\limits_{\mathbb{R}^2} \rmd^2 \bi{r}
\, u(\bi{r}) \, \overline{\varphi^{a, \beta, \bi{b}} (\bi{r})}.
\end{equation}
The integral for the coefficient $C_{\varphi}$ must be convergent.
This condition restricts the family of functions $\varphi$ what can
be used in these formulas.

Our idea is to take a solution of the wave equation and to use it as
a wavelet in the initial moment of time. Further the following
representation gives a solution of the wave equation:
\begin{eqnarray}
\widetilde{u}(\bi{r}, t) = \frac{1}{C_{\varphi}} \int\limits_{0}^{2\pi}
\rmd \beta \int\limits_{0}^{+\infty} \frac{\rmd a}{a^3} \,
\int\limits_{\mathbb{R}^2} \rmd^2\bi{b} \, U(a, \beta, \bi{b})
\widetilde{\varphi}^{a, \beta, \bi{b}}(\bi{r}, t), \nonumber \\
\widetilde{u}(\bi{r}, t)|_{t=0} = u(\bi{r}), \qquad
\widetilde{\varphi}^{a, \beta, \bi{b}}(\bi{r}, t)|_{t=0} =
\varphi^{a, \beta, \bi{b}}(\bi{r}),
\end{eqnarray}
and the family $\widetilde{\varphi}^{a, \beta, \bi{b}}$ is defined
by the formula
\begin{equation}
\widetilde{\varphi}^{a, \beta, \bi{b}}(\bi{r}, t) = \frac{1}{a}
\widetilde{\varphi}\left(\mathrm{M}^{-1}_{\beta}\, \frac{\bi{r} -
\bi{b}}{a}, \, \frac{t}{a} \right).
\end{equation}
The formulas written above are only an illustration of our method.
The question is whether a solution of any initial-value problem can
be represented in a similar form. We  want to present these
formulas in a common wavelet analysis formalism. There is also a problem of a convergence of integrals.

\section{Integral representation for solutions of the wave equation \label{S3}}

Generally speaking, we seek an integral representation of solutions
of the wave equation in the form
\begin{equation}\label{c3-General}
u(\bi{r}, t) = \int \rmd \mu(\nu) \,\, U(\nu) \,
\varphi^{\nu}(\bi{r}, t),
\end{equation}
where $\nu$ is a set of parameters, $\int \rmd \mu(\nu)$ denotes
integration with respect to the measure $\mu(\nu)$ in the space of
parameters, $\varphi^{\nu}(\bi{r}, t)$ is a family of elementary
solutions dependent on the parameter $\nu$, and the $U(\nu)$ are
coefficients. In the following sections, we define each of these
objects and  the Hilbert space $\mathcal{H}$ of solutions where such
a representation is allowed.

\subsection{The space $\mathcal{H}$ of solutions of the wave equation \label{S3-1}}

Consider the homogeneous wave equation in $\mathbb{R}^3$ with a
constant coefficient $c$:
\begin{equation}\label{c3-Wave-eq}
\Box \, u \equiv u_{tt} - c^2 \left( u_{xx} + u_{yy} + u_{zz}
\right) = 0, \qquad \bi{r} = (x,y,z).
\end{equation}
We fix the space $\mathcal{H}$ of complex-valued solutions of the
wave equation as a space of functions $u(\bi{r}, t)$, which are
square integrable with respect to the spatial coordinate $\bi{r}$
when the time $t$ is fixed and have the following Fourier transform
calculated with respect to the spatial coordinates when the time $t$
is fixed:
\begin{equation}
\widehat{u}(\bi{k}, t) = \widehat{u}_{+}(\bi{k}, 0) \, \rme^{-\rmi
|\bi{k}| ct} + \widehat{u}_{-}(\bi{k}, 0) \, \rme^{\rmi |\bi{k}|
ct}, \qquad \widehat{u}_{\pm}(\bi{k}, 0) \in L_{2}(\mathbb{R}^3).
\end{equation}
If the integrals
\begin{equation}\label{c3-class_cond}
\int\limits_{\mathbb{R}^3} \rmd^3 \bi{k} \,
|\widehat{u}_{\pm}(\bi{k}, 0)| < \infty, \qquad
\int\limits_{\mathbb{R}^3} \rmd^3 \bi{k} \, |\bi{k}|^2 \,
|\widehat{u}_{\pm}(\bi{k}, 0)| < \infty,
\end{equation}
converge, the solution $u(\bi{r}, t)$ allows one to take the second
derivative in the classical sense and the function $u(\bi{r}, t)$ is
a classical solution of (\ref{c3-Wave-eq}). If the integrals
(\ref{c3-class_cond}) diverge, this means that the solution
$u(\bi{r}, t)$ has discontinuities in the variable $\bi{r}$, and
thus is not a classical solution of the wave equation. In that case,
we introduce solutions in the sense of distributions. The function
$u(\bi{r}, t)$ is a solution of (\ref{c3-Wave-eq}) in the sense of
distributions \cite{Shilov-3} if it satisfies the equation
\begin{equation}\label{c3-wave-generalized}
\frac{\rmd^2}{\rmd t^2}\langle u(\bi{r}, t), \, \beta(\bi{r})
\rangle = c^2 \langle u(\bi{r}, t), \, \Delta \beta(\bi{r}) \rangle
\end{equation}
for all test functions $\beta(\bi{r})$ in a certain class. We
consider test functions $\beta(\bi{r})$ such that they decay at
infinity faster than $|\bi{r}|^m, \, \forall m
> 0$, and have all derivatives. This class of test functions is
usually named the Schwartz class $\mathbb{S}(\mathbb{R}^3)$ (see
\cite{Shilov-3}). The notation $\langle u(\bi{r}, t), \,
\beta(\bi{r}) \rangle$ stands for the common $L_2(\mathbb{R}^3)$
scalar product. If the function $u(\bi{r}, t) \in
L_{2}(\mathbb{R}^3)$ satisfies the equation
(\ref{c3-wave-generalized}) and have the second derivative in
$\bi{r}$ and $t$, it is a classical solution of the wave equation
(\ref{c3-Wave-eq}). We note here that even if the function
$u(\bi{r}, t)$ itself cannot be differentiated in $t$, it is easy to
show that the integral $\langle u(\bi{r}, t), \, \beta(\bi{r})
\rangle$ is a smooth function of $t$.

The space of solutions $\mathcal{H}$ is decomposed into a direct sum
of two subspaces $\mathcal{H}_{\pm}$ defined as follows:
\begin{equation}\label{c3-Hpm-def}
\mathcal{H} = \mathcal{H}_{+} \oplus \mathcal{H}_{-}, \qquad u =
u_{+} + u_{-},
\end{equation}
\begin{equation}
\mathcal{H}_{+} = \left\{u_{+}: \mathbb{R}^3 \times \mathbb{R}
\mapsto \mathbb{C} \,\, | \,\, u_{+} \in \mathcal{H}, \,
\widehat{u}_{+}(\bi{k}, t) = \widehat{u}_{+}(\bi{k}, 0)  \exp(-\rmi
|\bi{k}| c t) \right\},\nonumber
\end{equation}
\begin{equation}
\mathcal{H}_{-} = \left\{u_{-}: \mathbb{R}^3 \times \mathbb{R}
\mapsto \mathbb{C} \,\, | \,\, u_{-} \in \mathcal{H}, \,
\widehat{u}_{-}(\bi{k}, t) = \widehat{u}_{-}(\bi{k}, 0)  \exp(\rmi
|\bi{k}| c t) \right\}.\nonumber
\end{equation}
The spaces $\mathcal{H}_{\pm}$ consist of solutions supported on the
positive-frequency and negative-frequency light cones, respectively.

In the space $\mathcal{H}_{+}$, we introduce a common
$L_2(\mathbb{R}^3)$ scalar product with respect to the spatial
coordinates $\bi{r}$:
\begin{equation}
\langle u_{+}, \, v_{+} \rangle \equiv \int\limits_{\mathbb{R}^3}
\rmd^3\bi{r} \, u_{+}(\bi{r}, t) \, \overline{v_{+}(\bi{r}, t)} \, =
\, \frac{1}{(2\pi)^3} \int\limits_{\mathbb{R}^3} \rmd^3\bi{k} \,
\widehat{u}_{+}(\bi{k}, 0) \, \overline{\widehat{v}_{+}(\bi{k}, 0)}.
\end{equation}
This scalar product does not depend on the time $t$. This is the
main reason why we decompose the whole $\mathcal{H}$ into a direct
sum of these two subspaces $\mathcal{H}_{\pm}$. If we try to use the
common $L_2(\mathbb{R}^3)$ scalar product directly in the space
$\mathcal{H}$, the exponents $\exp(\rmi |\bi{k}|ct)$ and $\exp(-\rmi
|\bi{k}| ct)$ do not cancel, and the time dependance is not removed.
The case of $\mathcal{H}_{-}$ is analogous to $\mathcal{H}_{+}$.

\subsection{Wavelet-based integral representation for solutions from
$\mathcal{H}_{+}$ \label{sec-main}}

In this section, we give a decomposition of solutions from
$\mathcal{H}_{+}$ in terms of elementary solutions of the form
(\ref{c3-General}). First we determine a family of elementary
solutions of the wave equation (\ref{c3-Wave-eq}).

We fix a solution $\varphi_{+}(\bi{r}, t)$ of the wave equation
(\ref{c3-Wave-eq}) that belongs to the space $\mathcal{H}_{+}$. The
only constraint here is the following admissibility condition on the
Fourier transform $\widehat{\varphi}_{+}(\bi{k}, 0)$ of
$\varphi_{+}$:
\begin{equation}\label{c3a-Admis}
C_{\varphi}^{+} \equiv \int\limits_{\mathbb{R}^3} \rmd^3 \bi{k} \,\,
\frac{|\widehat{\varphi}_{+}(\bi{k}, 0)|^2}{|\bi{k}|^3} < \infty.
\end{equation}
We may call this solution a 'physical wavelet', following the
terminology introduced by G. Kaiser \cite{Kaiser-Book}. We also
assume that the physical wavelet belongs to $L_2(\mathbb{R}^3)
\bigcap L_1(\mathbb{R}^3)$ if the time $t$ is fixed. Then the
admissibility condition (\ref{c3a-Admis}) holds if the Fourier
transform $\widehat{\varphi}_{+}(\bi{k}, 0)$ has a root of any order
at the origin $\bi{k} = 0$. For the sake of simplicity, we assume
that the mother wavelet has an axial symmetry with respect to the
$OX$ axis. A more general case will also be considered below.

We construct a family of elementary solutions in the following way.
We apply translations by $\bi{b} \in \mathbb{R}^3$, dilations by $a
> 0$, and rotations through angles $\vartheta_1, \,
\vartheta_2$ to the spatial coordinates $\bi{r}$. We also introduce
the dilation of the time $t$ by $a$. We denote by
$\varphi^{\nu}_{+}(\bi{r}, t)$ the family of solutions obtained,
supplying it with the superscript $\nu$:
\begin{equation}\label{c3a-Family}
\varphi_{+}^{\nu}(\bi{r}, t) \equiv \frac{1}{a^{3/2}}
\varphi_{+}\left(\mathrm{M}^{-1}_{\vartheta_1
\vartheta_2}\frac{\bi{r} - \bi{b}}{a}, \, \frac{t}{a} \right),
\qquad \nu = (a, \bi{b}, \vartheta_1, \vartheta_2),
\end{equation}
where $\mathrm{M}_{\vartheta_1, \vartheta_2}$ is defined as
follows:
\begin{eqnarray}\label{c2-matrix}
 \mathrm{M}_{\vartheta_1 \vartheta_2} = \left(\begin{array}{ccc}
\cos\vartheta_1 & -\sin\vartheta_1 & 0 \\
\sin\vartheta_1 & \cos\vartheta_1 & 0 \\
0 & 0 & 1
\end{array}\right) \,
\left(\begin{array}{ccc}
1 & 0 & 0 \\
0 & \cos\vartheta_2 & -\sin\vartheta_2 \\
0 & \sin\vartheta_2 & \cos\vartheta_2
\end{array} \right) .
\end{eqnarray}
In this notation, the three parameters $a, \vartheta_1, \vartheta_2$
have the meaning of a spatial frequency vector $\bi{q}$, where
$|\bi{q}|$ is proportional to $1/a$ and the angles define its direction in spherical
coordinate system.

Now we define the coefficients $U(\nu)$ of the decomposition
(\ref{c3-General}). We suggest to put these coefficients equal to
the  wavelet transform of the solution $u_{+}(\bi{r}, t) \in
\mathcal{H}_{+}$. Then the coefficients $U_{+}(\nu)$ are defined in
terms of the scalar product of $u_{+}$ and $\varphi_{+}^{\nu}$:
\begin{equation}\label{c3a-Coefs}
U_{+}(\nu) \equiv \langle u_{+}, \, \varphi_{+}^{\nu} \rangle =
\frac{1}{a^{3/2}} \int\limits_{\mathbb{R}^3} \rmd^3\bi{r} \,
u_{+}(\bi{r}, t) \, \overline{\varphi_{+} \left(
\mathrm{M}^{-1}_{\vartheta_1, \vartheta_2}\frac{\bi{r}
- \bi{b}}{a}, \, \frac{t}{a} \right)} \nonumber
\end{equation}
\begin{equation}
= \frac{a^{3/2}}{(2\pi)^3} \int\limits_{\mathbb{R}^3} \rmd^3 \bi{k}
\,\, \widehat{u}_{+}(\bi{r}, 0) \, \exp(\rmi \bi{k} \cdot \bi{b}) \,
\overline{\widehat{\varphi}_{+}\left(a\mathrm{M}^{-1}_{\vartheta_1,
\vartheta_2}\bi{k}, \, 0\right)}.
\end{equation}
Since the scalar product in $\mathcal{H}_{\pm}$ does not depend on
$t$, the coefficients $U_{+}(\nu)$ do not depend on time as well.

Below we will use a fact common to wavelet analysis techniques. The
wavelet transform of any square integrable functions possesses the
isometry property \cite{Daubechies, Antoine-Book}
\begin{equation}\label{Isometry-known}
\langle f(\bi{r}), \, g(\bi{r}) \rangle = \frac{1}{C_{\varphi}} \int
\rmd \mu(\nu) \,  F(\nu) \, \overline{G(\nu)}.
\end{equation}
Here $f, g$ are some square integrable functions, and $F, G$ are
their wavelet transforms. If functions are solutions of the wave
equation, they depend on the time $t$ as a parameter.

We apply (\ref{Isometry-known}) for two solutions $u_+(\bi{r},t),$
$v_+(\bi{r},t)$ taken at one and the same fixed moment. We can write
\begin{eqnarray}\label{c3a-Isometry}
\langle u_{+}, \, v_{+} \rangle = \frac{1}{C^{+}_{\varphi}} \int
\rmd \mu(\nu) \,  U_{+}(\nu) \, \overline{V_{+}(\nu)}, \qquad
\forall u_{+}, \, v_{+} \in \mathcal{H}_{+}, \\
\int \rmd\mu(\nu) = \int\limits_{0}^{2\pi} \rmd \vartheta_1
\int\limits_{0}^{\pi} \rmd \vartheta_2 \, \sin\vartheta_2
 \, \int\limits_0^{\infty}
\frac{\rmd a}{a^4} \int\limits_{\mathbb{R}^3} \rmd^3 \bi{b}.
\label{c3a-dmu}
\end{eqnarray}
This property extends to any other moment of time, because neither
$U_{+}(\nu)$, $V_{+}(\nu)$ nor the scalar product of $u_{+}$ and
$v_{+}$ depend on the time $t$. The isometry property implies the
reconstruction formula
\begin{equation}\label{c3a-Reconstr}
u_{+}(\bi{r}, t) = \frac{1}{C_{\varphi}^{+}} \int \rmd\mu(\nu) \,
U_{+}(\nu) \, \varphi_{+}^{\nu}(\bi{r}, t),
\end{equation}
which holds in the weak sense for any $u_{+} \in \mathcal{H}_{+}$.
The coefficients $U_{+}(\nu)$ do not depend on $\bi{r}$ and $t$ and
thus the formula (\ref{c3a-Reconstr}) has the meaning of a
superposition of elementary solutions $\varphi_{+}^{\nu}(\bi{r},
t)$. Similar arguments can be applied to the space
$\mathcal{H}_{-}$. The weak sense in the last formula means that we
can take the $L_2$ inner product of $u_{+}$ and any solution $v_{+}
\in \mathcal{H}_{+}$ on the left-hand side and the inner product of
$\varphi^{\nu}_{+}$ and $v_{+}$ on the right-hand side under the
sign of integration and obtain equality. This formula can also be
interpreted from the point of view of distributions of four
variables (see \ref{ap1}).

Below we show that the relation (\ref{c3a-Reconstr}) is valid not
only if we take the scalar product with a solution $v_{+}(\bi{r},
t)$, but also if we use a test function $\beta(\bi{r})$ from
$\mathbb{S}(\mathbb{R}^3)$ instead of $v_{+}$.  This means that we
can understand the relation (\ref{c3a-Reconstr}) in the sense of
distributions:
\begin{equation}\label{distr-short}
\langle u_{+}(\bi{r}, t), \beta(\bi{r})\rangle =
\frac{1}{C_{\varphi}^{+}} \int \rmd\mu(\nu) \, U_{+}(\nu) \,
\langle\varphi_{+}^{\nu}(\bi{r}, t),\beta(\bi{r})\rangle,
\end{equation}
where the function $\beta(\bi{r})$ is an arbitrary test function
defined above.

The proof of (\ref{distr-short}) follows the line of argument given
below. First we note that the inner product can be written in the
form
\begin{equation}\label{int}
\langle u_{+}(\bi{r},t), \, \beta(\bi{r})\rangle =
\int\limits_{\mathbb{R}^3} \,\rmd^3 \bi{r} \,u_+(\bi{r},t) \,
\overline{\beta(\bi{r})} \nonumber
\end{equation}
\begin{equation}
= \frac{1}{(2\pi)^3} \int\limits_{\mathbb{R}^3} \,\rmd^3 \bi{k}
\,\widehat{u}_+(\bi{k},0)\, \rme^{-\rmi \omega_{+}(\bi{k})\, t} \,
\overline{\widehat{\beta}(\bi{k})} = \langle u_{+}(\bi{r}, 0), \,
\beta_{-}(\bi{r},t)\rangle,
\end{equation}
where
\begin{equation}
\beta_{-}(\bi{r},t) = \frac{1}{(2\pi)^3} \int\limits_{\mathbb{R}^3}
\rmd^3 \bi{k}\, \widehat{\beta}(\bi{k})\, \rme^{\rmi\bi{k}
\cdot\bi{r} + \rmi \omega_{+}(\bi{k})\, t}.
\end{equation}
The standard isometry formula  (\ref{Isometry-known}) gives
\begin{equation}\label{short1}
\langle u_{+}(\bi{r}, 0), \beta_{-}(\bi{r},t)\rangle =
\frac{1}{C^{+}_{\varphi}} \int \rmd \mu(\nu) \,  U_{+}(\nu) \,
\overline{B_{-}(\nu, t)},
\end{equation}
where the wavelet transform of the function $\beta_{-}(\bi{r},t)$
can be written as
\begin{equation}\label{short2}
B_{-}(\nu, t) \equiv  \langle \beta_{-}(\bi{r}, t), \,
\varphi_{+}^{\nu}(\bi{r}, 0) \rangle \nonumber
\end{equation}
\begin{equation}
= \langle \beta(\bi{r}), \, \varphi_{+}^{\nu}(\bi{r}, t) \rangle =
\overline{\langle \varphi_{+}^{\nu}(\bi{r}, t) , \, \beta(\bi{r})
\rangle}.
\end{equation}
Substituting (\ref{short2}) into (\ref{short1}), we obtain
(\ref{distr-short}).

Moreover a relation similar to (\ref{distr-short}) but containing
the derivatives in time is valid:
\begin{equation}\label{distr-short-der}
\frac{\mathrm{d}^k \langle u_{+}(\bi{r}, t), \beta(\bi{r})\rangle
}{\mathrm{d} t^k} = \frac{1}{C_{\varphi}^{+}} \int \rmd\mu(\nu) \,
U_{+}(\nu) \, \frac{\mathrm{d}^k \langle\varphi_{+}^{\nu}(\bi{r},
t),\beta(\bi{r})\rangle}{\mathrm{d} t^k},
\end{equation}
where $k$ is any integer. We take the relation (\ref{int}) as a
starting point and differentiate it under the sign of integral $k$
times. For $\beta \in \mathbb{S}$ we obtain
\begin{equation}
\frac{\mathrm{d}^k \langle u_{+}(\bi{r}, t), \beta(\bi{r})\rangle
}{\mathrm{d} t^k} = \left\langle u_{+}(\bi{r}, 0),
\frac{\mathrm{d}^k \beta_{-}(\bi{r},t)}{\mathrm{d} t^k}
\right\rangle.
\end{equation}
The isometry property (\ref{Isometry-known}) results in
\begin{equation}\label{short1-d}
\left\langle u_{+}(\bi{r}, 0), \frac{\mathrm{d}^k
\beta_{-}(\bi{r},t)} {\mathrm{d} t^k} \right\rangle =
\frac{1}{C^{+}_{\varphi}} \int \rmd \mu(\nu) \,  U_{+}(\nu) \,
\overline{B_{-}^{(k)}(\nu, t)},
\end{equation}
where the wavelet transform $B_{-}^{(k)}$ can be written in the form
\begin{equation}\label{short2-d}
B_{-}^{(k)}(\nu, t) \equiv  \left\langle \frac{\mathrm{d}^k
\beta_{-}(\bi{r}, t)}{\mathrm{d} t^k}, \, \varphi_{+}^{\nu}(\bi{r},
0) \right\rangle \nonumber
\end{equation}
\begin{equation}
= \frac{\mathrm{d}^k \langle \beta_{-}(\bi{r},t), \,
\varphi_{+}^{\nu}(\bi{r}, 0) \rangle}{\mathrm{d} t^k} =
\frac{\mathrm{d}^k \langle \beta(\bi{r}), \,
\varphi_{+}^{\nu}(\bi{r}, t) \rangle}{\mathrm{d} t^k}.
\end{equation}
Substituting (\ref{short2-d}) into (\ref{short1-d}), we obtain
(\ref{distr-short-der}). This means that if the function $\varphi$
is a solution of (\ref{distr-short}), then the integral is also a
solution in the sense of distributions.

\subsection{Simplifications and generalizations of the integral formula for solutions}
The number of parameters in the family of solutions can be reduced and the formulas
(\ref{c3a-Family}) - (\ref{c3a-Reconstr}) can be simplified if the
mother wavelet $\varphi_{+}$ is a spherically symmetric one. Then
the set of parameters reduces to $\nu = (a, \bi{b})$. The
admissibility condition (\ref{c3a-Admis}) takes a simpler form:
\begin{equation}\label{c3a-Admis-Sph}
C^{+}_{\varphi} \equiv 4\pi \int\limits_{0}^{\infty} \rmd k \,
\frac{ |\widehat{\varphi}_{+}(k, 0)|^2 }{k} < \infty, \qquad k =
|\bi{k}|.
\end{equation}
The family of wavelets reads
\begin{equation}\label{c3a-Family-Sph}
\varphi_{+}^{a, \bi{b}}(\bi{r}, t) \equiv \frac{1}{a^{3/2}}
\varphi_{+}\left(\frac{\bi{r} - \bi{b}}{a}, \, \frac{t}{a} \right),
\end{equation}
and the representation has the form
\begin{equation}\label{c3a-Reconstr-Sph}
u_{+}(\bi{r}, t) = \frac{4\pi}{C^{+}_{\varphi}}
\int\limits_{0}^{\infty} \frac{\rmd a}{a^4}
\int\limits_{\mathbb{R}^3} \rmd^3\bi{b} \, U_{+}(a, \bi{b}) \,
\varphi_{+}^{a, \bi{b}}(\bi{r}, t).
\end{equation}
Although this particular case of analysis is the simplest one, it is
blind to directional properties of wave fields, and this may be
inconvenient for practical purposes.

The most general case occurs when the mother wavelet $\varphi$
possesses no symmetry at all. Then we should use the full set of
three Euler angles $\vartheta_1, \vartheta_2$ and $\vartheta_3$ to
determine rotations of the mother wavelet in a three-dimensional
space. The set of parameters now reads $\widetilde{\nu} = (a,
\bi{b}, \vartheta_1, \vartheta_2, \vartheta_3)$. The rotation matrix
now has the form
\begin{equation}\label{c2-matrix-Eul}
\mathrm{M}_{\vartheta_1 \vartheta_2 \vartheta_3}
\end{equation}
\begin{equation}
= \left(\begin{array}{ccc}
\cos\vartheta_1 & -\sin\vartheta_1 & 0 \\
\sin\vartheta_1 & \cos\vartheta_1 & 0 \\
0 & 0 & 1
\end{array}\right) \,
\left(\begin{array}{ccc}
1 & 0 & 0 \\
0 & \cos\vartheta_2 & -\sin\vartheta_2 \\
0 & \sin\vartheta_2 & \cos\vartheta_2
\end{array} \right) \,
\left(\begin{array}{ccc}
\cos\vartheta_3 & -\sin\vartheta_3 & 0 \\
\sin\vartheta_3 & \cos\vartheta_3 & 0\\
0 & 0 & 1
\end{array} \right),\nonumber
\end{equation}
and the reconstruction formula reads
\begin{equation}\label{c3a-reconstr-all}
 u_{+}(\bi{r}, t) = \frac{1}{C^{+}_{\varphi}}
\int\limits_{0}^{2\pi} \rmd \vartheta_1 \int\limits_{0}^{\pi} \rmd
\vartheta_2 \, \sin\vartheta_2 \int\limits_{0}^{2\pi} \rmd
\vartheta_3 \, \int\limits_0^{\infty} \frac{\rmd a}{a^4}
\int\limits_{\mathbb{R}^3} \rmd^3 \bi{b} \, U_{+}(\widetilde{\nu}) \,
\varphi_{+}^{\widetilde{\nu}}(\bi{r}, t).
\end{equation}

Another generalization of the reconstruction formulas is associated
with the possibility of using different mother wavelets in
calculating wavelet transforms $U_{+}(\nu)$, $V_{+}(\nu)$ (see
\cite{Daubechies}, \cite{Antoine-Book}). Let $\psi_{+}(\bi{r}, t),$
$\chi_{+}(\bi{r}, t) \in \mathcal{H}_{+}$. Upon constructing
families of wavelets $\psi_{+}^{\nu},$ $\chi_{+}^{\nu}$ similarly to
(\ref{c3a-Family}), we calculate wavelet transforms in the form
\begin{equation}\label{Coefs}
U_{+}(\nu) \equiv \langle u_{+}, \, \psi_{+}^{\nu} \rangle, \quad
V_{+}(\nu) \equiv \langle v_{+}, \, \chi_{+}^{\nu} \rangle.
\end{equation}
The isometry property is valid in this case as well, but instead of
the constant $C^{+}_{\varphi}$ in the formula (\ref{c3a-Isometry})
we must take the constant
\begin{equation}\label{con-psi-chi}
C_{\psi \chi}^{+} \equiv \int\limits_{\mathbb{R}^3} \rmd^3 \bi{k}
\,\, \frac{\overline{\widehat{\psi}_{+}(\bi{k},
0)}\widehat{\chi}_{+}(\bi{k}, 0) }{|\bi{k}|^3}.
\end{equation}
Then we obtain the reconstruction formula
\begin{equation}\label{c3a-Reconstr-2}
u_{+}(\bi{r}, t) = \frac{1}{C_{\psi \chi}^{+}} \;\int \rmd\mu(\nu) \,
U_{+}(\nu) \, \chi_{+}^{\nu}(\bi{r}, t),
\end{equation}
which holds at least in the weak sense and in the sense of
distributions. If $t=0$, then the formula (\ref{c3a-Reconstr-2}) is
a well-known wavelet reconstruction formula (see \cite{Daubechies},
\cite{Antoine-Book}). The formulas (\ref{c3a-Reconstr-Sph}) and
(\ref{c3a-reconstr-all}) can be generalized in a similar way.

\subsection{Initial-value problem for the wave equation}

Since each solution of the wave equation from $\mathcal{H}_{\pm}$
can easily be expressed in terms of its initial-value problem, it is
useful to obtain formulas for the solution of the initial-value
problem in terms of localized solutions. These formulas will allow
one to avoid splitting the solution $u$ into its positive-frequency
and negative-frequency parts $u_{+}$ and $u_{-}$, respectively, in
order to calculate the coefficients of the decomposition. The
splitting requires an additional operation - taking the Fourier
transform of the solution, which is not convenient in some cases.
Consider the following initial-value problem:
\begin{equation}\label{c5-Cauchy}
\left\{ \begin{array}{l} u_{tt} - c^2(u_{xx} + u_{yy} + u_{zz}) = 0,
\\ \displaystyle{u|_{t=0} } =  w(\bi{r}), \qquad
\displaystyle{\left. \frac{\partial u}{\partial t} \right|_{t = 0} }
= v(\bi{r}), \end{array} \right.
\end{equation}
\begin{equation}
w(\bi{r}) \in L_2(\mathbb{R}^3), \,\,\, v(\bi{r}) \in
L_2(\mathbb{R}^3). \nonumber
\end{equation}
We seek a solution of the form
\begin{equation}\label{u+u-}
u(\bi{r}, t) = \frac{1}{C_{\varphi}^{+}} \int \rmd\mu(\nu) \,
U_{+}(\nu) \, \varphi_{+}^{\nu}(\bi{r}, t) +
\frac{1}{C_{\varphi}^{-}} \int \rmd\mu(\nu) \, U_{-}(\nu) \,
\varphi_{-}^{\nu}(\bi{r}, t).
\end{equation}
In this section, we construct the solution formally. We will justify
this solution in the sense of distributions in the next subsection.

We need to find the decomposition coefficient $U_{\pm}(\nu)$. The
expression (\ref{c3a-Coefs}) for $U_{\pm}$ requires $u_{+}$ and
$u_{-}$ or their Fourier transforms. We obtain $\widehat{u}_{\pm}$
from the initial-value data. To find an idea for this, we use a
well-known Fourier formula for a solution of this initial-value
problem. It reads
\begin{equation}\label{c5-Cauchy-Four}
 u(\bi{r}, t) = \frac{1}{2(2\pi)^3} \int\limits_{\mathbb{R}^3}
\rmd^3\bi{k} \left[\widehat{w}(\bi{k}) - \frac{1}{\rmi c |\bi{k}| }
\widehat{v}(\bi{k}) \right] \, \exp(\rmi\bi{k} \cdot \bi{r} - \rmi
|\bi{k}| c t) \nonumber
\end{equation}
\begin{equation} + \frac{1}{2(2\pi)^3}
\int\limits_{\mathbb{R}^3} \rmd^3\bi{k} \left[\widehat{w}(\bi{k}) +
\frac{1}{\rmi c |\bi{k}| } \widehat{v}(\bi{k}) \right] \,
\exp(\rmi\bi{k} \cdot \bi{r} + \rmi |\bi{k}| c t).
\end{equation}
This formula yields
\begin{equation}\label{c5-upm}
\widehat{u}_{+}(\bi{k}, 0) = \frac{1}{2}\left[\widehat{w}(\bi{k}) -
\frac{1}{\rmi c |\bi{k}| } \widehat{v}(\bi{k})\right], \,\,
\widehat{u}_{-}(\bi{k}, 0) = \frac{1}{2}\left[\widehat{w}(\bi{k}) +
\frac{1}{\rmi c |\bi{k}| } \widehat{v}(\bi{k})\right].
\end{equation}
Substituting $\widehat{u}_{\pm}$ from (\ref{c5-upm}) into
(\ref{c3a-Coefs}), we obtain
\begin{equation}
U_{\pm}(\nu) = \frac{1}{2(2\pi)^3} \int\limits_{\mathbb{R}^3} \rmd^3
\bi{k} \,\, \widehat{w}(\bi{k}) \,
\overline{\widehat{\varphi}_{\pm}^{\nu}(\bi{k}, \, 0)} \,\,
\nonumber
\end{equation}
\begin{equation}
\mp \,\, \frac{1}{2(2\pi)^3} \int\limits_{\mathbb{R}^3} \rmd^3
\bi{k} \,\, \frac{1}{\rmi c|\bi{k}|} \, \widehat{v}(\bi{k}) \,
\overline{\widehat{\varphi}_{\pm}^{\nu}(\bi{k}, \, 0)}.
\label{U-trans}
\end{equation}
We introduce solutions $\psi_{\pm} \in \mathcal{H}_{\pm}$ of the
wave equation in such a way that
\begin{equation}\label{phi-psi}
\varphi_{\pm}(\bi{r}, \, t) = \pm \frac{\partial}{\partial
t}\psi_{\pm}(\bi{r}, t), \quad \widehat{\varphi}_{\pm}(\bi{k}, 0) =
- \rmi\,c\,|\bi{k}|\, \widehat{\psi}_{\pm}(\bi{k}, 0).
\end{equation}
Families of solutions are built in the standard way:
\begin{equation}\label{f-Fam}
\psi_{\pm}^{\nu}(\bi{r}, t) \equiv \frac{1}{a^{3/2}}
\psi_{\pm}\left(\mathrm{M}^{-1}_{\vartheta_1
\vartheta_2}\frac{\bi{r} - \bi{b}}{a}, \, \frac{t}{a} \right),
\qquad \nu = (a, \bi{b}, \vartheta_1, \vartheta_2),
\end{equation}
then
\begin{equation}
\varphi_{\pm}^{\nu}(\bi{r}, t) = \pm a \frac{\partial}{\partial
t}\psi_{\pm}^{\nu}(\bi{r}, t), \qquad
\widehat{\varphi}^{\nu}_{\pm}(\bi{k}, 0) =  - \rmi\,c\,a\,|\bi{k}|
\,\, \widehat{\psi}^{\nu}_{\pm}(\bi{k}, 0).
\end{equation}
The formula (\ref{U-trans}) in the new notation reads
\begin{equation}
U_{\pm}(\nu) = \frac{1}{2} \int\limits_{\mathbb{R}^3} \rmd^3 \bi{r}
\,\, w(\bi{r}) \, \overline{\varphi_{\pm}^{\nu} (\bi{r}, \, 0)} \mp
\frac{a}{2} \int\limits_{\mathbb{R}^3} \rmd^3 \bi{r} \,\, v(\bi{r})
\, \overline{\psi_{\pm}^{\nu} (\bi{r}, \, 0)} \nonumber
\end{equation}
\begin{equation}
= \frac{1}{2} W_{\pm}(\nu) \mp \frac{a}{2} V_{\pm}(\nu),
\end{equation}
where the capital letters $W_{\pm}$ and $V_{\pm}$ denote the wavelet
transforms of $w$ and $v$ with respect to wavelets $\varphi_{\pm}$
and $\psi_{\pm}$, respectively:
\begin{eqnarray}
W_{\pm}(\nu) = \langle w(\bi{r}), \varphi_{\pm}^{\nu}(\bi{r},
0)\rangle, \qquad V_{\pm}(\nu) = \langle v(\bi{r}),
\psi_{\pm}^{\nu}(\bi{r}, 0)\rangle.
\end{eqnarray}
Now the reconstruction formula of the form (\ref{u+u-}) for the sum
of $u_{+}$ and $u_{-}$  reads:
\begin{equation}
u(\bi{r}, t) = \frac{1}{C^{+}_{\varphi}} \int\rmd \mu(\nu) \,\,
\left[ \frac{1}{2} W_{+}(\nu) - \frac{a}{2} V_{+}(\nu) \right] \,
\varphi_{+}^{\nu}(\bi{r}, \, t) \nonumber
\end{equation}
\begin{equation}
+ \,\, \frac{1}{C^{-}_{\varphi}} \int\rmd \mu(\nu) \,\, \left[
\frac{1}{2} W_{-}(\nu) + \frac{a}{2} V_{-}(\nu) \right] \,
\varphi_{-}^{\nu}(\bi{r}, \, t). \label{c5-Rec1}
\end{equation}

\subsection{Justification of results}

In this subsection, we show that the formal solution (\ref{c5-Rec1})
can be understood at least in the sense of distribution. This means
that we understand the wave equation (\ref{c5-Cauchy}) in the sense
(\ref{distr-short}) and we state the initial conditions in the form
\begin{eqnarray}\label{Cauchy-distr}
  \langle u(\bi{r}, 0), \beta(\bi{r}) \rangle  = \langle
w(\bi{r}), \beta(\bi{r}) \rangle, \qquad \left. \frac{\rmd \langle
u(\bi{r}, t), \beta(\bi{r}) \rangle }{\rmd t} \right|_{t = 0} =
\langle v(\bi{r}), \beta(\bi{r}) \rangle.
\end{eqnarray}
We use the fact that $u(\bi{r}, t)\to u(\bi{r}, 0)$ in the $L_2$
sense as $t \to 0$. This allows us to write $\langle u(\bi{r}, 0),
\beta(\bi{r}) \rangle$ instead of $\langle u(\bi{r}, t),
\beta(\bi{r}) \rangle|_{t=0}$. However we cannot write $\langle
\partial u(\bi{r}, t)/\partial t, \beta(\bi{r}) \rangle$ in the
sense of a scalar product in $L_2$, because the function $\partial
u(\bi{r}, t)/\partial t$ in general does not belong to $L_2$ and
belongs to the class of distributions. Instead, we use the fact that
the scalar product $\langle u(\bi{r}, t), \beta(\bi{r}) \rangle$ can
be differentiated with respect to $t$ even if the solution
$u(\bi{r}, t)$ itself is not differentiable (see \cite{Shilov-3}).
This allows us to pose correctly the initial value problem even for
nonclassical solutions. If the functions $\varphi_{\pm}$ are
solutions of the wave equation in the sense of (\ref{distr-short}),
the integral in (\ref{c5-Rec1}) is also a solution of the wave
equation in the sense of distributions, as was shown in Section
\ref{sec-main}. Now we show that $u(\bi{r}, t)$ defined by
(\ref{c5-Rec1}) also satisfies the initial conditions
(\ref{Cauchy-distr}).

It is useful to take into consideration the auxiliary solution
$\chi_{\pm}(\bi{r}, t) \in \mathcal{H}_{\pm},$
\begin{equation}\label{chi-phi}
\chi_{\pm}(\bi{r}, \, t) = \pm \frac{\partial}{\partial
t}\varphi_{\pm}(\bi{r}, t), \qquad \widehat{\chi}_{\pm}(\bi{k}, 0) =
- \rmi\,c\,|\bi{k}|\, \widehat{\varphi}_{\pm}(\bi{k}, 0).
\end{equation}
The family of solutions $\chi_{\pm}^{\nu}$, constructed by analogy
with (\ref{f-Fam}), satisfies the relations
\begin{equation}\label{chi-def}
\chi_{\pm}^{\nu}(\bi{r}, t) = \pm a \frac{\partial}{\partial
t}\varphi_{\pm}^{\nu}(\bi{r}, t), \qquad
\widehat{\chi}^{\nu}_{\pm}(\bi{k}, 0) =  - \rmi\,c\,a\,|\bi{k}| \,\,
\widehat{\varphi}^{\nu}_{\pm}(\bi{k}, 0).
\end{equation}
First we consider the simplest case where $\varphi_{-}(\bi{r}, t) =
\varphi_{+}(\bi{r}, -t),$ $\varphi_{+}(\bi{r}, t) \equiv
\varphi(\bi{r}, t)$. Then $C_{\varphi} \equiv C_{\varphi}^{+} =
C_{\varphi}^{-},$ $W_+ = W_- \equiv W,$ $V_+ = V_- \equiv V.$ We fix the time $t$ equal to zero and, upon
regrouping the terms, obtain
\begin{equation}
\langle u(\bi{r}, 0), \beta(\bi{r}) \rangle   =
\frac{1}{2C_{\varphi}} \int\rmd \mu(\nu) \,\, \left[ W(\nu) + W(\nu)
\right] \, \langle \varphi^{\nu}(\bi{r}, \, 0), \beta(\bi{r})
\rangle   \nonumber
\end{equation}
\begin{equation}
+ \, \frac{1}{2C_{\varphi}} \int\rmd \mu(\nu) \, \left[ - a V(\nu) + a
V(\nu) \right] \, \langle \varphi^{\nu}(\bi{r}, \, 0), \beta(\bi{r})
\rangle   \, = \, \langle w(\bi{r}), \beta(\bi{r})\rangle.
\end{equation}
Here we used the isometry property (\ref{Isometry-known}) for
$w(\bi{r})$ and $\beta(\bi{r})$. Next we consider the time
derivative at the moment of time $t = 0$. We note that
$\varphi^{\nu}_t(\bi{r}, t)|_{t=0} = -\varphi^{\nu}_t (\bi{r},
-t)|_{t=0}$. Upon regrouping the terms, we obtain
\begin{equation}
\left. \frac{\partial}{\partial t} \langle u(\bi{r}, t),
\beta(\bi{r}) \rangle \right|_{t=0} = \frac{1}{2C_{\varphi}}
\int\rmd \mu(\nu) \,\, W(\nu) \left[ \left. \frac{\partial}{\partial
t} \langle \varphi^{\nu}(\bi{r}, t), \beta(\bi{r}) \rangle
\right|_{t=0} \right. \nonumber
\end{equation}
\begin{equation}
\left. + \left.\frac{\partial}{\partial t} \langle
\varphi^{\nu}(\bi{r}, -t), \beta(\bi{r}) \rangle \right|_{t=0}
\right] - \, \frac{1}{2C_{\varphi}} \int\rmd \mu(\nu) \,\, a V(\nu)
\left[ \left. \frac{\partial}{\partial t} \langle
\varphi^{\nu}(\bi{r}, t), \beta(\bi{r}) \rangle \right|_{t=0}
\right. \nonumber
\end{equation}
\begin{equation}
\left. - \left. \frac{\partial}{\partial t}
\langle\varphi^{\nu}(\bi{r}, -t), \beta(\bi{r}) \rangle
\right|_{t=0} \right] = - \frac{1}{C_{\varphi}} \int\rmd \mu(\nu) \,\,
a V(\nu) \left. \frac{\partial}{\partial t} \langle
\varphi^{\nu}(\bi{r}, t), \beta(\bi{r}) \rangle \right|_{t=0}
\nonumber
\end{equation}
\begin{equation}
= \frac{1}{C_{\psi\chi}} \int\rmd \mu(\nu) \,\,
V(\nu)\langle\chi^{\nu}(\bi{r}, 0), \beta(\bi{r}) \rangle.
\label{dut}
\end{equation}
Here we use the fact that
\begin{equation}\nonumber
\left.\frac{\rmd}{\rmd t} \left\langle \varphi^{\nu}(\bi{r}, t),
\beta(\bi{r}) \right\rangle \right|_{t=0} = \left.\left\langle
\frac{\partial}{\partial t} \varphi^{\nu}(\bi{r}, t), \beta(\bi{r})
\right\rangle \right|_{t=0} = \langle \chi^{\nu}(\bi{r}, 0),
\beta(\bi{r}) \rangle,
\end{equation}
which follows from the Plancherel equality and the fact that
${\partial}\varphi^{\nu}/{\partial t} \equiv \chi^{\nu}$ is square
integrable in $\bi{r}$. The last-mentioned integral gives us the
function $\langle v(\bi{r}), \beta(\bi{r}) \rangle$ in accordance
with the formulas (\ref{c3a-Reconstr-2}). The constant $C_{\psi
\chi}$ can be calculated, using (\ref{phi-psi}) and
(\ref{chi-phi}); $C_{\psi \chi} = - C_{\varphi}$.

Now we consider the general case where $\varphi_{+}(\bi{r}, 0) \neq
\varphi_{-}(\bi{r}, 0)$. We understand all the relations below in
the sense of (\ref{distr-short}) and (\ref{distr-short-der}) and
their analogs in $\mathcal{H}_{-}$.  In contrast to the previous
case, we cannot regroup terms in such a way that the expressions
under the sign of integration cancel. We must show instead that the
integrals cancel. First for the time $t = 0$ the integrals in
(\ref{c5-Rec1}) multiplied by $\beta(\bi{r})$ containing terms
$W_{+}(\nu) / 2$ and $W_{-}(\nu) / 2$ yield the product $\langle
w(\bi{r}),{\beta}(\bi{r})\rangle$ by a wavelet inverse transform
formula (the formula (\ref{c3a-Reconstr-2}) for $t=0$).  We show
that other two integrals containing $aV_{+}(\nu)\langle
\varphi_+^{\nu}(\bi{r}, 0),{\beta}(\bi{r})\rangle/2$ and
$-aV_{-}(\nu)\langle \varphi_-^{\nu}(\bi{r},
0),{\beta}(\bi{r})\rangle/2$ cancel and provide the zero sum. Taking
into account the formulas (\ref{chi-def}), we obtain
\begin{equation}
a\langle \varphi_+^{\nu}(\bi{r},0),\beta(\bi{r})\rangle =
\frac{a}{(2\pi)^3} \int\limits_{\mathbb{R}^3} \rmd^3 \bi{k} \,
\overline{\widehat{\beta}(\bi{k})} \,
\widehat{\varphi}_{+}^{\nu}(\bi{k}, 0) \,\, \nonumber
\end{equation}
\begin{equation}
\,\, = \,\, \frac{1}{(2\pi)^3} \int\limits_{\mathbb{R}^3} \rmd^3
\bi{k} \, \overline{\left( \frac{a \widehat{\beta}(\bi{k})}{\rmi
\,c\, a|\bi{k}|}\right) }\, \widehat{\chi}_{+}^{\nu}(\bi{k}, 0) \,\,
= \,\, \overline{\widetilde{B}_{+}(\nu)}, \label{app01}
\end{equation}
where $\widetilde{B}_{+}(\nu)$ is the wavelet transform of a certain
function $\widetilde{\beta}(\bi{r})$ that has the Fourier transform
of the form $\widehat{\beta}(\bi{k}) / \rmi c|\bi{k}|$:
\begin{equation}
\widetilde{\beta}(\bi{r}) = \frac{1}{(2\pi)^3}
\int\limits_{\mathbb{R}^3} \rmd^3\bi{k} \;\;\frac{
\widehat{\beta}(\bi{k})}{\rmi \,c\,|\bi{k}|} \, \exp{(- \rmi \;
\bi{k} \cdot \bi{r} )}.
\end{equation}
The singularity at the point $\bi{k}=0$ is integrable because the
function $\beta(\bi{r})\in \mathbb{S}$, and thus
$\widehat{\beta}(\bi{k})$ is continuous. On applying the isometry
relation to the integral containing $aV_{+}(\nu)/2$ in
(\ref{c5-Rec1}), we obtain
\begin{equation}
\frac{1}{C^{+}_{\varphi}} \int\rmd \mu(\nu) \,\,  \frac{a}{2}
V_{+}(\nu)  \, \langle\varphi_{+}^{\nu}(\bi{r}, \, 0),
\beta(\bi{r})\rangle =- \frac{1}{2} \langle v(\bi{r},0),
\,\widetilde{\beta}(\bi{r})\rangle.
\end{equation}
Here we use the fact that $C^{+}_{\psi\chi} = -C^{+}_{\varphi}$ by
the definitions of $\psi$ and $\chi$ in (\ref{phi-psi}) and
(\ref{chi-phi}). The integral containing $aV_{-}(\nu)/2$ gives us
the same term but with an opposite sign. Their sum is equal to zero;
then the formula (\ref{c5-Rec1}) for $t = 0$ provides $\langle
w(\bi{r}), \beta(\bi{r})\rangle$.

Now we calculate the time derivative of the expression
(\ref{c5-Rec1}) multiplied by $\beta(\bi{r})$ at the moment of time
$t = 0$. Here the sum of terms containing $aV_{+}(\nu)/2$ and
$-aV_{-}(\nu)/2$ yields $\langle v(\bi{r}), \beta(\bi{r}) \rangle$
for the same reasons as those concerning the formula (\ref{dut}).
For example, in the positive-frequency case we have
\begin{equation}
\frac{1}{C_{\varphi}^+} \int\rmd \mu(\nu) \,\, a V_+(\nu) \left.
\left\langle \frac{\partial}{\partial t} \varphi_+^{\nu}(\bi{r}, t)
\right|_{t=0}, \beta(\bi{r}) \right\rangle \nonumber
\end{equation}
\begin{equation}
= \frac{1}{C^+_{\varphi}} \int\rmd \mu(\nu) \,\,  V_+(\nu)
\langle\chi_{+}^{\nu}(\bi{r}, 0), \beta(\bi{r}) \rangle = \langle
v(\bi{r}), \beta(\bi{r}) \rangle.
\end{equation}
Next we show that the sum of integrals containing $W_{\pm}(\nu)/2$
is equal to zero:
\begin{equation}
\frac{1}{2C_{\varphi}^{+}}\int\rmd \mu(\nu) \, W_{+}(\nu) \left.
\left\langle \frac{\partial}{\partial t} \varphi_{+}^{\nu}(\bi{r},
t) \right|_{t = 0}, \beta(\bi{r}) \right\rangle \,\,  \nonumber
\end{equation}
\begin{equation}
= \,\,\frac{1}{2C_{\varphi}^{+}}\int\rmd \mu(\nu) \, W_{+}(\nu)
\frac{\langle \chi_+^{\nu}(\bi{r}, 0), \beta(\bi{r}) \rangle}{a}.
\nonumber
\end{equation}
By analogy with (\ref{app01}), we have
\begin{equation}
\frac{\langle \chi_+^{\nu}(\bi{r}, 0), \beta(\bi{r}) \rangle}{a} = -
\frac{1}{a(2\pi)^3} \int\limits_{\mathbb{R}^3} \rmd^3 \bi{k} \, \rmi
c a |\bi{k}| \, \widehat{\varphi}_{+}^{\nu}(\bi{k}, 0)
\,\overline{\widehat{\beta}(\bi{k})} \,\, \nonumber
\end{equation}
\begin{equation}
= \,\, \frac{1}{(2\pi)^3} \int\limits_{\mathbb{R}^3} \rmd^3 \bi{k}
\, \widehat{\varphi}_{+}^{\nu}(\bi{k}, 0)\, \overline{{\rmi \,c\,
|\bi{k}|} \, \widehat{\beta}(\bi{k})} \,\, =
\,\,\overline{{\breve{B}}_{+}(\nu)},
\end{equation}
where $\breve{B}_{+}(\nu)$ is the wavelet transform of a function
$\breve{\beta}(\bi{r})$ that has the Fourier transform of the form
$\widehat{\beta}(\bi{k}) \; \rmi c|\bi{k}|$:
\begin{equation}
\breve{\beta}(\bi{r}) = \frac{1}{(2\pi)^3}
\int\limits_{\mathbb{R}^3} \rmd^3\bi{k} \;\;{
\widehat{\beta}(\bi{k})}\;{\rmi \,c\,|\bi{k}|} \, \exp{(- \rmi \;
\bi{k} \cdot \bi{r} )}.
\end{equation}
The integral containing $W_{+}(\nu)$ in (\ref{c5-Rec1}) multiplied
by $\beta$ is equal to half the scalar product $\langle w(\bi{r}),
\breve{\beta}(\bi{r})\rangle$. The integral containing $W_{-}(\nu)$
and multiplied by $\beta$ gives the same term $\langle w(\bi{r}),
\breve{\beta}(\bi{r})\rangle/2$ but with an opposite sign. Then
their sum is equal to zero and the time derivative of the expression
(\ref{c5-Rec1}) multiplied by $\beta$ taken at the moment $t = 0$ is
equal to the function $\langle v(\bi{r}), \beta(\bi{r})\rangle$.
Then the formula (\ref{c5-Rec1}) actually yields a solution of the
initial-value problem (\ref{c5-Cauchy}). Since each solution of the
wave equation from $\mathcal{H}$ can be represented in terms of its
initial-value problem data, the integral representation is valid for
any solution $u \in \mathcal{H}$.

\subsection{Comparison with the results of Kaiser}

We compare our formulas (\ref{c3a-Family}) -
(\ref{c3a-reconstr-all}) with the results obtained by G. Kaiser in
\cite{Kaiser-Book}. He defines the coefficients of the decomposition
in terms of the analytic-signal transform (AST) (\ref{c3-Kaiser-AS})
of the solution under consideration. Further the AST of the solution
is regarded in a spatial frequency domain and is interpreted as the
scalar product (\ref{c3-Kaiser-AS2}) of this solution and the
wavelet (\ref{c3-Kaiser-Wavelet}). Therefore the formula for this
wavelet is strictly determined by the expression for the AST. The
parameter $\bi{x} \in \mathbb{R}^3$ of the AST (\ref{c3-Kaiser-AS})
has the meaning of translation and the parameter $s \in \mathbb{R},
\, s \neq 0$, has the meaning of dilation. No rotation is used,
owing to the spherical symmetry of the wavelet
(\ref{c4-Kaiser-wavelet}), (\ref{c3-Kaiser-Wavelet}). The
representation formula (\ref{c3-Kaiser-Repr}) obtained by Kaiser
then coincides, up to notation and normalization, with our special
formula (\ref{c3a-Reconstr-Sph}). It should also be noted that
Kaiser uses another norm (\ref{c3-Kaiser-NORM}) and do not decompose
the whole space $\mathcal{H}$ into $\mathcal{H}_{\pm}$ explicitly.

The main difference between our and Kaiser's approaches is that we
start our definitions with the decomposition of the space
$\mathcal{H}$ into $\mathcal{H}_{+}$ and $\mathcal{H}_{-}$ and the
introduction of time-independent scalar products in each of them,
instead of using AST. This allows us to choose a mother wavelet from
the wide class (\ref{c3a-Admis}) rather than the from fixed one
(\ref{c3-Kaiser-Wavelet}) used by G. Kaiser \cite{Kaiser-Book}.

\section{Some examples of physical wavelets \label{S4}}

As is clearly seen from Section \ref{S3-1}, we can construct the
physical wavelet for $\mathcal{H}_{\pm}$ just by choosing an
arbitrary square integrable function
$\widehat{\varphi}_{\pm}(\bi{k})$ of three variables $\bi{k} \in
\mathbb{R}^3$ having a root of any order at the point $\bi{k} = 0$,
then by multiplying it by the time-depending exponent $\exp(\mp \rmi
|\bi{k}| c t)$, and by taking the Fourier inverse transform with
respect to the spatial frequency coordinates $\bi{k}$. This provides
admissible physical wavelets in $\mathcal{H}_{\pm}$. However, in
practice we possibly will be unable to take the integral
analytically and find an exact formula for the wavelet $\varphi$ in
the position space. There are several practical methods that allow
one to obtain an exact solution of the wave equation directly in the
position space, without integrating over the whole $\mathbb{R}^3$.
These methods were originated in papers
\cite{Bateman}-\cite{BesShaZiol} (see \cite{Kiselev-Review} for a
review of such methods). The aim of this section is to look at some
of these methods from the point of view of physical wavelets. We
find conditions for these solutions to be admissible physical
wavelets.

\subsection{Spherically symmetric mother wavelets \label{S4-1}}

G. Kaiser uses in \cite{Kaiser-Book} the following method for
constructing his physical wavelet. He considers two solutions of the
inhomogeneous wave equation, namely, the emitted wave
$\texttt{e}(\bi{r}, t)$ and the absorbed wave $\texttt{a}(\bi{r},
t)$:
\begin{eqnarray}\label{c4-e1}
\left(\frac{\partial^2}{\partial t^2} - c^2 \Delta \right)
\texttt{e}(\bi{r}, t) = \phi(ct) \, \delta(\bi{r}), \qquad
\texttt{e}(\bi{r}, t) = \frac{1}{4\pi c^2} \frac{ \phi(ct -
|\bi{r}|)}{|\bi{r}|}, \\ \left(\frac{\partial^2}{\partial t^2} - c^2
\Delta \right) \texttt{a}(\bi{r}, t) = \phi(ct) \, \delta(\bi{r}),
\qquad \texttt{a}(\bi{r}, t) = \frac{1}{4\pi c^2} \frac{\phi(ct +
|\bi{r}|)}{|\bi{r}|},\label{c4-e2}
\end{eqnarray}
where $\phi$ is a function of the time $t$. These solutions are
spherically symmetric and have singularities at the origin $\bi{r} =
0$. The solution $\texttt{e}(\bi{r}, t)$ has the meaning of a wave
emitted by a point source at the origin, and the solution
$\texttt{a}(\bi{r}, t)$ has the meaning of a solution absorbed by a
point source. The difference between these two functions
\begin{equation}\label{c4-sph}
\varphi(\bi{r}, t) = \texttt{a}(\bi{r},t) - \texttt{e}(\bi{r},t) =
\frac{1}{4\pi c^2 |\bi{r}|} \left[ \phi(ct + |\bi{r}|) - \phi(ct -
|\bi{r}|) \right]
\end{equation}
is a solution of the homogeneous wave equation (\ref{c3-Wave-eq}).
Upon subtraction, the singularity at the origin $\bi{r} = 0$
cancels. The function $\phi(ct)$ was called by G. Kaiser in
\cite{Kaiser-Book} a '\textit{proxy wavelet}'.

We find conditions that should be applied to the class of proxy
wavelets $\phi$ in order to obtain admissible physical wavelets. The
Fourier transform of the solution $\varphi(\bi{r}, t)$
(\ref{c4-sph}) can be calculated exactly. It reads
\begin{eqnarray}\label{c4-Fourier-Of-Sph}
\widehat{\varphi}(\bi{k}, t) = - \frac{\widehat{\phi}( |\bi{k}|) }{2
\rmi |\bi{k}| c^2} \exp( \rmi  |\bi{k}| c t) +
\frac{\widehat{\phi}(-|\bi{k}|) }{2 \rmi |\bi{k}| c^2} \exp(- \rmi
|\bi{k}| c t),
\end{eqnarray}
where $\widehat{\phi}(\pm |\bi{k}|)$ is the Fourier transform of the
function $\phi(t)$ taken at the point $\pm |\bi{k}|$. This formula
shows that the Fourier transform $\widehat{\varphi}(\bi{k}, t)$
splits into two terms, depending on $\exp( -\rmi |\bi{k}| c t)$ and
$\exp( \rmi |\bi{k}| c t).$

We stress here that if we choose a progressive proxy wavelet
$\phi(t)$, i.e., $\widehat{\phi}(\xi) \equiv 0$ for $\xi < 0$, the
wavelet $\varphi$ constructed will belong to $\mathcal{H}_{-}$ only,
and it can be marked with subscript $-$, i.e. $\varphi(\bi{r}, t) =
\varphi_{-}(\bi{r}, t)$. The second wavelet $\varphi_{+} \in
\mathcal{H}_{+}$ in that case can be obtained from $\varphi_{-}$ by
changing the sign of the time variable $t$. The admissibility
condition (\ref{c3a-Admis-Sph}) then can be stated in terms of the
proxy wavelet $\phi$ in the way
\begin{equation}
C^{-}_{\varphi} = \frac{\pi}{c^4} \int\limits_{0}^{+\infty}  \rmd k
\, \frac{|\widehat{\phi}(k)|^2}{k^3} \,\, < \,\, \infty.
\end{equation}

The physical wavelet (\ref{c3-Kaiser-Wavelet}) introduced by Kaiser
in \cite{Kaiser-Book} was interpreted also as a solution derived
from the following proxy wavelet:
\begin{equation}\label{c4-k1}
\phi(t) = \frac{\Gamma(\alpha)}{ \pi(1 -  \rmi t)^{\alpha}}, \qquad
\widehat{\phi}(\xi) = 2\Theta(\xi) \, \xi^{\alpha - 1} \exp(-\xi),
\end{equation}
where $\Theta$ is the Heaviside step function. In the position
space, this wavelet has the form
\begin{equation}
\varphi_{+}(\bi{r}, t) = \frac{\Gamma(\alpha)}{4\pi^2 c^2 |\bi{r}|}
\left[ \frac{1}{[1 -  \rmi  (ct + |\bi{r}|)]^{\alpha}} - \frac{1}{[1
-  \rmi ( ct - |\bi{r}|)]^{\alpha}} \right].
\label{c4-Kaiser-wavelet}
\end{equation}
Since the proxy wavelet $\phi(t)$ can be chosen from a wide class of
functions, we can obtain other spherically symmetric physical
wavelets with better properties from the point of view of wavelet
analysis. To show this, we provide here another example of
spherically symmetric physical wavelet, which has, in contrast to
that  suggested by G. Kaiser, an exponential decay and an
infinite number of zero moments. This example of a solution first
appeared in \cite{Iwo}, and in \cite{Perel-Sidorenko-2} was first
regarded as a mother wavelet. It can also be interpreted in terms
of a field of two point sources (\ref{c4-e1}), (\ref{c4-e2}) with
the proxy wavelet
\begin{equation}
\phi(t) = \exp\left(-2 \sqrt{1 - \rmi t} \right).
\end{equation}
The branch of the square root with positive real part is implied
here and below. The difference between the absorbed and emitted
waves can be taken as a mother wavelet in $\mathcal{H}_{-}$ and has
the form
\begin{equation}
\varphi_{-}(\bi{r}, t) = \frac{1}{4\pi c^2 |\bi{r}| }
\left\{\exp\left[-2\sqrt{1 - \rmi (ct + |\bi{r}|)}\right] \right.
\nonumber
\end{equation}
\begin{equation}
\left. -\exp\left[-2\sqrt{1 - \rmi (ct - |\bi{r}| )}\right]
\right\}. \label{c4-Our-Wavelet-Sph}
\end{equation}
The Fourier transform $\widehat{\varphi}_{-}(\bi{k}, t)$ reads
\begin{equation}\label{c4-Our-Wavelet-Fourier-Sph}
\widehat{\varphi}_{-}(\bi{k}, 0) = \frac{\rmi\sqrt{\pi}}{c^2} \,\,
|\bi{k}|^{-5/2} \, \exp\left[- |\bi{k}| - \frac{1}{|\bi{k}|}\right].
\end{equation}
The coefficient $C^{-}_{\varphi}$ (\ref{c3a-Admis-Sph}) for the
wavelet $\varphi_{-}$ can be calculated exactly:
\begin{equation}
C^{-}_{\varphi} = 4\pi \int\limits_{0}^{\infty}  \rmd  k \,
\frac{|\widehat{\varphi}_{-}(k, 0)|^2}{k} \,\, = \,\,
\frac{8\pi^2}{c^4}\, K_{5} (4) \,\, < \,\, \infty,
\end{equation}
where $K_{5}(4)$ is a McDonald's function \cite{Abramovitz-Stegan}
of order $5$. The Fourier transform of this wavelet has a root of
infinite order at the origin $\bi{k} = 0$ owing to the factor
$\exp(-1 / |\bi{k}|)$ and the wavelet itself has an infinite number
of zero moments. The wavelet $\varphi_{-}$ has a spherical symmetry
and an exponential decay away from the circle where $|\bi{r}| = ct$.

\subsection{Nonsymmetric mother wavelets \label{S4-2}}

We discuss here the construction of nonsymmetric solutions of the
wave equation (\ref{c3-Wave-eq}) following papers \cite{Ziolkowski,
BesShaZiol}. The method is based on the summation of well-known
nonstationary Gaussian beams \cite{Brittingham, Kiselev} multiplied
by a weight function $\widehat{\phi}(q)$:
\begin{equation}\label{c4-Beam-Summation}
\varphi(\bi{r}, t) = \int\limits_{0}^{+\infty}  \rmd q \,
\widehat{\phi}(q) \, \varphi_{\mathrm{beam}}(q, \bi{r}, t),
\end{equation}
\begin{equation}
\label{c4-Gaus-beam} \varphi_{\mathrm{beam}}(q, \bi{r}, t)  =
\frac{\exp{[\rmi \, q \, \theta(\bi{r}, t)  ]} }{ \sqrt{x + ct -
\rmi \varepsilon_1} \sqrt{x + ct - \rmi \varepsilon_2}}, \qquad q \,
> \,0,
\end{equation}
\begin{equation}
\theta(\bi{r}, t) = x - ct + \frac{y^2}{x + ct - \rmi \varepsilon_1}
+ \frac{z^2}{x + ct - \rmi \varepsilon_2}, \nonumber
\end{equation}
where $\varepsilon_{1}$ and $\varepsilon_2$ are free positive
parameters. If $\widehat{\phi}(q) \equiv 0, \, q < 0$, the formula
(\ref{c4-Beam-Summation}) has the meaning of a Fourier inverse
transform and can be written in a simpler form
\begin{eqnarray}\label{c4-Bateman}
\varphi(\bi{r}, t) = \frac{1}{ \sqrt{x + ct - \rmi \varepsilon_1}
\sqrt{x + ct - \rmi \varepsilon_2}} \, \phi[\theta(\bi{r}, t)],
\end{eqnarray}
where $\phi$ is a Fourier inverse transform of $\widehat{\phi}$. The
function $\phi$ then can be named a 'proxy wavelet' following the
terminology of G. Kaiser. The formula (\ref{c4-Bateman}) is a
special case of the class of solutions presented by Bateman in
\cite{Bateman, Bateman-Book} and further developed by Hillion in
\cite{Hillion}. Now we determine the class of proxy wavelets $\phi$
that produce admissible physical wavelets. The Fourier transform of
a solution $\varphi(\bi{r}, t)$ defined by (\ref{c4-Bateman}) reads
\begin{equation}\label{c4-Four-Batemann}
\widehat{\varphi}(\bi{k}, t) = 2\pi^2 \rmi \widehat{\phi} \left(
\frac{ k_{\mathrm{x}} + |\bi{k}| }{2} \right) \frac{1}{|\bi{k}|} \,
\exp\left[ - \rmi |\bi{k}| c t - \frac{k^2_{\mathrm{y}}
\varepsilon_1 + k^2_{\mathrm{z}} \varepsilon_2 }{2 (k_{\mathrm{x}} +
|\bi{k}|) } \right],
\end{equation}
\begin{equation}
\bi{k} = (k_{\mathrm{x}}, k_{\mathrm{y}}, k_{\mathrm{z}}). \nonumber
\end{equation}
Substituting this expression into the formula for the coefficient
$C_{\varphi}$ defined by (\ref{c3a-Admis}), we conclude that
$\widehat{\phi}(q)$ must have a root of order at least $1 + \alpha,
\,\, \alpha > 0$, at the origin $q = 0$. We also restrict the class
of admissible proxy wavelet to the class $L_1(\mathbb{R}) \bigcap
L_2(\mathbb{R})$.

A special case of solutions of the class (\ref{c4-Bateman}) named
the Gaussian wave packet was found in \cite{Kiselev-Perel-OptSp,
Kiselev-Perel-JMP} and studied in \cite{Perel-Sidorenko-JPA}:
\begin{equation}\label{c4-Gaus-Pack}
\varphi(\bi{r}, \, t) = \frac{1}{   \sqrt{x + ct - \rmi
\varepsilon_1} \sqrt{x + ct - \rmi \varepsilon_2} } \exp\left[ -p
\sqrt{1 - \frac{\rmi\theta(\bi{r}, t)}{\gamma}} \right],
\end{equation}
where $p$ and $\gamma$ are free positive parameters. This solution
can be obtained from (\ref{c4-Bateman}) by using the following proxy
wavelet:
\begin{equation}\label{c4-Our-Proxy}
\phi(t) = \exp\left[ -p \sqrt{1 - \frac{\rmi t}{\gamma}} \right].
\end{equation}
The Fourier transform of the Gaussian packet (\ref{c4-Gaus-Pack})
due to (\ref{c4-Four-Batemann}) and (\ref{c4-Our-Proxy}) has the
form
\begin{equation}
\widehat{\varphi}(\bi{k}, \, t) = \rmi(2\pi)^{3/2}
\frac{p}{\sqrt{\gamma}} \, \frac{1}{|\bi{k}| (|\bi{k}| +
k_\mathrm{x})^{3/2}} \nonumber
\end{equation}
\begin{equation}
\times \exp\left[ -\frac{|\bi{k}| + k_\mathrm{x}}{2}\gamma -
\frac{p^2}{2\gamma} \frac{1}{|\bi{k}| + k_\mathrm{x}} -
\frac{k^2_{\mathrm{y}} \varepsilon_1 + k^2_{\mathrm{z}}
\varepsilon_2 }{2 (k_{\mathrm{x}} + |\bi{k}|) }  - \rmi
|\bi{k}|ct\right].
\end{equation}
This physical wavelet has an exponential decay away from the moving
point $x = ct, y = 0, z = 0$. It has infinitely many zero moments
with respect to spatial coordinates. As is shown in
\cite{Perel-Sidorenko-JPA}, its asymptotics coincides with the
Morlet wavelet \cite{Daubechies, Antoine-Book} as $p \to \infty$ and
the time $t$ is fixed:
\begin{equation}
\varphi(\bi{r}, t) =  \frac{C}{(-\rmi \varepsilon_1)^{1/2} (-\rmi
\varepsilon_2)^{1/2} } \exp{\left[ \rmi \varkappa (x - c t) - \frac{
(x - c t)^2}{2\sigma_\mathrm{x}^2} -
\frac{y^2}{2\sigma_{\mathrm{y}}^2} -
\frac{z^2}{2\sigma_{\mathrm{z}}^2} \right] } \nonumber
\end{equation}
\begin{equation}
\times {\left[1 + \Or( p^{-3\alpha+1}) \right]}, \qquad \alpha \in
(1/3, \, 1/2),
\end{equation}
where
\begin{equation}
\sigma_\mathrm{x}^2 = { 4\gamma^2 }/{p }, \,\,\,
\sigma_{\mathrm{y}}^2 = \gamma \varepsilon_1/p, \,\,\,
\sigma_{\mathrm{z}}^2 = \gamma \varepsilon_2/p, \qquad \varkappa =
\frac{p}{2\gamma},
\end{equation}
in the domain
\begin{eqnarray}\label{c4-vicinity}
(x-ct) / \gamma = \Or \left(p^{-\alpha}\right), \quad y /
\sqrt{\varepsilon_1 \gamma}  = \Or \left(p^{-\alpha}\right), \quad z
/ \sqrt{\varepsilon_2 \gamma}  = \Or \left(-p^{\alpha}\right),
\end{eqnarray}
provided that the parameters $2ct / \varepsilon_j, \,\,
p^{-\alpha}\gamma / \varepsilon_j, \,\, j = 1,2,$ are small.

The axially symmetric case of the Gaussian beam (\ref{c4-Gaus-beam})
and the Gaussian packet (\ref{c4-Gaus-Pack}) can be obtained by
putting $\varepsilon_1 = \varepsilon_2$.

\section{Conclusions}

A new integral representation of solutions of the wave equation was
built. It is based on mathematical methods of a continuous wavelet
transform in a three-dimensional space. An arbitrary solution of the
wave equation can be represented as a superposition of elementary
solutions. We discussed methods of constructing these elementary
solutions.  Both spherically symmetric and axially symmetric
elementary solutions were considered. Examples of known physical
wavelets were given. Solution of the initial-value problem based on
wavelet analysis was also considered. A brief comparison of the
presented results with the results obtained by G. Kaiser was also
carried out. The integral representation constructed in this article
may be useful in real problems of wave propagation.

\section*{Acknowledgments}
M.Sidorenko was supported by the Dmitry Zimin 'DYNASTY' Foundation
\\

\appendix

\section{The representation build by G. Kaiser \label{ap0}}

The first integral representation of the form (\ref{c3-General})
based on mathematical methods of wavelet analysis was presented by
G. Kaiser in his book \cite{Kaiser-Book}. We give here a brief
review of his results for the scalar wave equation (or acoustic
equation). The results for vector Maxwell equations were also
presented in \cite{Kaiser-Book}. In this Appendix we follow the
specific notation introduced by Kaiser.

A solution of the wave equation can be represented in the following
form by means of the Fourier transform:
\begin{eqnarray}\label{Kaiser-Fourier}
F(\mathrm{x}) = \int\limits_{\mathbb{R}^3} \frac{\rmd^3 \bi{p}}{16
\pi^3 \omega} \, \left[\rme^{\rmi(\omega t - \bi{p} \cdot \bi{x})}
\, f(\bi{p}, \omega)  + \rme^{\rmi(-\omega t - \bi{p} \cdot \bi{x})}
\,
f(\bi{p}, -\omega) \right], \\
\mathrm{x} = (\bi{x}, t), \,\,\,\, \bi{x} \in \mathbb{R}^3, \qquad
\mathrm{p} = (\bi{p}, p_0), \,\,\,\, \bi{p} \in \mathbb{R}^3, \qquad
\omega = |\bi{p}|. \nonumber
\end{eqnarray}
Here, in accordance with the notation by G. Kaiser, $F(\mathrm{x}),
\,\, \mathrm{x} = (\bi{x}, t)$ denotes a solution of the wave
equation and $f(\mathrm{p}), \,\, \mathrm{p} = (\bi{p}, p_0)$, is
the Fourier transform of $F(\mathrm{x})$ taken with respect to both
spatial and time coordinates, using the Lorentz-covariant scalar
product $\mathrm{p} \cdot \mathrm{x} = p_0 t - \bi{p} \cdot \bi{x}$.
Owing to the wave equation, the function $f(\mathrm{p})$ depends not
on four variables $\bi{p}, p_0$ but on three variables $\bi{p}$ in
the following two ways: $f(\bi{p}, |\bi{p}|)$ or $f(\bi{p},
-|\bi{p}|)$.

The norm of the solution $F$ is defined in the way
\begin{equation}\label{c3-Kaiser-NORM}
\| F \|^2 = \int\limits_{\mathbb{R}^3} \frac{\rmd^3 \bi{p}}{16 \pi^3
\omega^{\alpha}} \, \left[|f(\bi{p}, \omega)|^2 + |f(\bi{p},
-\omega)|^2\right], \qquad \alpha > 2.
\end{equation}
The analytic signal transform of $F(\mathrm{x})$ plays the role of
coefficients for the decomposition (\ref{c3-General}):
\begin{eqnarray}\label{c3-Kaiser-AS}
\widetilde{F}(\mathrm{x} + \rmi \mathrm{y}) \equiv \frac{1}{\pi
\rmi} \int\limits_{-\infty}^{\infty} \frac{\rmd \tau}{\tau - \rmi}
\, F(\mathrm{x} + \tau \mathrm{y}), \qquad \mathrm{x} = (\bi{x}, t),
\,\,\,\, \mathrm{y} = (\bi{y}, y_0).
\end{eqnarray}
Kaiser showed that it is sufficient to put $\mathrm{z} = \mathrm{x}
+ \rmi \mathrm{y}$ equal to $(\bi{x}, \rmi s), \,\, \bi{x} \in
\mathbb{R}^3, \, s \neq 0$. Then the analytic signal transform
(\ref{c3-Kaiser-AS}) takes a simpler form in the Fourier domain:
\begin{eqnarray}\label{c3-Kaiser-AS2}
 \widetilde{F}(\bi{x}, \rmi s) = \int\limits_{\mathbb{R}^3}
\frac{\rmd^3 \bi{p}}{8 \pi^3 \omega^{\alpha}} \, \omega^{\alpha - 1}
\, \rme^{-\rmi \bi{p} \cdot \bi{x}} \, \left[\Theta(s) \rme^{-\omega
s} \, f(\bi{p}, \omega) + \Theta(-s) \rme^{\omega s} \, f(\bi{p},
-\omega) \right],
\end{eqnarray}
where $\Theta$ is the Heaviside step function. Further an arbitrary
solution $F(\mathrm{x})$ with the finite norm (\ref{c3-Kaiser-NORM})
can be represented in the form
\begin{equation}
F(\mathrm{x}') = \int\limits_{E} \rmd \mu_{\alpha}(\mathrm{z}) \,
\Psi_{\mathrm{z}}(\mathrm{x}') \, \widetilde{F}(\mathrm{z}), \qquad
\mathrm{x}' = (\bi{x}', t), \,\,\, \mathrm{z} = (\bi{x}, \rmi s),
\label{c3-Kaiser-Repr}
\end{equation}
\begin{equation}
\rmd \mu_{\alpha}(\mathrm{z}) = \frac{2^{\alpha - 3}}{\Gamma(\alpha
- 2)} \, \rmd^3 \bi{x}\, \rmd s \, |s|^{\alpha - 3}, \qquad E = \{
(\bi{x}, \rmi s), \,\, \bi{x} \in \mathbb{R}^3, \, s \neq 0\},
\nonumber
\end{equation}
\begin{equation}
\Psi_{\bi{x}, \rmi s}(\mathrm{x}') = \int\limits_{\mathbb{R}^3}
\frac{\rmd^3 \bi{p}}{8 \pi^3 \omega} \, \omega^{\alpha - 1} \,
\rme^{\rmi \bi{p} \, \cdot (\bi{x} - \bi{x}') } \,
\left[\Theta(\omega s)\, \rme^{-\omega s + \rmi \omega t} +
\Theta(-\omega s) \, \rme^{\omega s - \rmi \omega t} \right].
\label{c3-Kaiser-Wavelet}
\end{equation}
The integral (\ref{c3-Kaiser-Repr}) is a special case of the
representation in the most general form (\ref{c3-General}). The
coefficients $\widetilde{F}(\mathrm{z})$ do not depend on time and
coordinates, and thus the formula (\ref{c3-Kaiser-Repr}) has the
meaning of a superposition of solutions (\ref{c3-Kaiser-Wavelet}).
The solution (\ref{c3-Kaiser-Wavelet}) is also considered in Section
\ref{S4-1} in connection with proxy wavelets. In the position space
it is given by the formula (\ref{c4-Kaiser-wavelet}).

A disadvantage of the formula (\ref{c3-Kaiser-Repr}) is that we can
use only one solution $\Psi(\mathrm{x}')$ as a 'building block' for
the reconstruction. The method presented in this paper is free of
this disadvantage.

\section{Relationship between the  scalar product in ${\cal H}$ and
four-dimensional distributions \label{ap1}}

In Section \ref{S3-1}, we introduced a solution of the wave equation
in the sense of distributions. We regarded solutions as
distributions in three spatial variables $\bi{r} = (x,y,z)$ and the
time $t$ that is viewed as the parameter. This approach is similar
to that presented in \cite{Shilov-3}. However we may view solutions
as distributions in all variables $\bi{r}, t$. Then we must use
four-variable test functions $\alpha(\bi{r}, t) \in
\mathbb{S}(\mathbb{R}^4)$. The Fourier transform of a solution $u
\in \mathcal{H}$ calculated in four variables has the form
\begin{equation}
F[u](\bi{k}, \omega) = \delta(\omega + c |\bi{k}|)
\;\widehat{u}_{+}(\bi{k}, 0) + \delta(\omega - c |\bi{k}|)\;
\widehat{u}_{-}(\bi{k}, 0).
\end{equation}
Next we calculate the action of $u$ on $\alpha(\bi{r}, t) \in
\mathbb{S}(\mathbb{R}^4)$, using the Plancherel equality $(u(\bi{r},
t), \, \alpha(\bi{r}, t) ) = 1/2\pi^3 \, ( \widehat{u}(\bi{k},
\omega), \, \widehat{\alpha}(\bi{k}, \omega) )$. We obtain
\begin{equation}\label{u-f-u}
( u(\bi{r}, t), \alpha(\bi{r}, t)) = \left\langle
\,\widehat{u}_{+}(\bi{k}, 0), \, \overline{\alpha(\bi{k},
c|\bi{k}|)}\, \right\rangle + \left\langle \,\widehat{u}_{-}(\bi{k},
0), \,\overline{ \alpha(\bi{k}, -c|\bi{k}|)}\, \right\rangle.
\end{equation}
We introduce solutions $A_+(\bi{r}, t)$ and $A_-(\bi{r}, t)$ by
means of the inverse Fourier transforms of $\overline{\alpha(\bi{k},
c|\bi{k}|)}$ and $\overline{\alpha(\bi{k}, -c|\bi{k}|)}$,
respectively, in the form
\begin{equation}
 A_+ = \frac{1}{(2\pi)^3} \int\limits_{\mathbb{R}^3} \rmd^3\bi{k}
\, \overline{\alpha(\bi{k}, c|\bi{k}|)}\, \rme^{\rmi\bi{k} \cdot
\bi{r} - \rmi\omega t},\;\; A_- = \frac{1}{(2\pi)^3}
\int\limits_{\mathbb{R}^3} \rmd^3\bi{k}  \,\overline{\alpha(\bi{k},
-c|\bi{k}|)}\, \rme^{\rmi\bi{k} \cdot \bi{r}+ \rmi\omega
t}.\nonumber
\end{equation}
Finally, formula (\ref{u-f-u}) can be written in terms of these
solutions:
\begin{equation}\label{u-f-u-ok}
( u(\bi{r}, t), \alpha(\bi{r}, t) ) = \langle {u}_{+}(\bi{r}, t), \,
A_+(\bi{r}, t)\rangle + \langle  {u}_{-}(\bi{r}, t), \, A_-(\bi{r},
t) \rangle.
\end{equation}
We emphasize that $\alpha(\bi{r}, t)$ is not a solution of the wave
equation, but $A_+(\bi{r}, t)$, $A_-(\bi{r}, t)$, which are
constructed by means of $\alpha,$ are solutions. So we conclude that
the action of $u$ on $\alpha \in \mathbb{S}(\mathbb{R}^4)$ can be
expressed in terms of the scalar product in $\mathcal{H}$.

\end{document}